\RequirePackage{color}
\documentclass{PoS}

\title{Assessing Theory Uncertainties in EFT Power Countings from Residual Cutoff Dependence}

\ShortTitle{Assessing Theory Uncertainties from Residual
  Cutoff Dependence}

\author{\speaker{Harald W.~Grie{\ss}hammer}%\thanks{}
  \\Institute for Nuclear Studies, Department of Physics, George
  Washington University, Washington DC 20052, USA; E-mail: \email{hgrie@gwu.edu}}

\abstract{I summarise a method to quantitatively assess the consistency of
  power-counting proposals in Effective Field Theories which are
  non-perturbative at leading order. It uses the fact that the Renormalisation
  Group evolution of an observable predicts the functional form of its
  residual cutoff dependence on the EFT breakdown scale, on the low-momentum
  scales, and on the order of the calculation.  Passing this test is a
  necessary but not sufficient consistency criterion for a suggested power
  counting whose exact nature is disputed.  For example, in \ChiEFT with more
  than one nucleon, a lack of universally accepted analytic solutions
  obfuscates the relation between convergence pattern and numerical results,
  and led to proposals which predict different numbers of Low Energy
  Coefficients at the same chiral order. The method may provide an independent
  check whether an observable is properly renormalised at a given
  order, and allows one to estimate both the breakdown scale and the
  momentum-dependent order-by-order convergence pattern of an EFT. Conversely,
  it may help identify those LECs which produce renormalised observables at a
  given order. I also discuss its underlying assumptions and relation to the
  Wilsonian Renormalisation Group Equation; useful choices for observables and
  cutoffs; the momentum window in which the test provides best signals; its
  dependence on the values and forms of cutoffs as well as on the EFT
  parameters; the impact of fitting Low Energy Coefficients to data in
  different or the same channel; and caveats as well as limitations. Since the
  test is designed to minimise the use of data, it allows one to
  quantitatively falsify if the EFT has been renormalised consistently, rather
  than quantifying how an EFT compares to experiment. Its application in
  particular to the \wave{3}{P}{0} and \wave{3}{P}{2}-\wave{3}{F}{2} partial
  waves of $\mathrm{NN}$ scattering in \ChiEFT may elucidate persistent
  power-counting issues. Details and a better bibliography can be found in an
  upcoming publication~\cite{hgrie}.}

\FullConference{The 8th International Workshop on Chiral Dynamics, CD2015 ***\\
		29 June 2015 - 03 July 2015\\
		Pisa,Italy}

   \usepackage{grffile} % allows graphics files to have "." in filename
\usepackage{slashed}
\usepackage{multirow}                   %%% tabs over multiple rows
\usepackage{graphicx} % for figures
\usepackage{bm}       % bold math 
\usepackage{amssymb}
\usepackage{amsmath}
\usepackage{xspace}                    %%% correct spacing self-defined macros

\newcommand{\ii}{\mathrm{i}}
\newcommand{\dd}{\mathrm{d}}

\newcommand{\half}{\frac{1}{2}}

\newcommand{\mpi}{\ensuremath{m_\pi}}     % pion mass
\newcommand{\MeV}{\ensuremath{\mathrm{MeV}}}

\newcommand{\ChiEFT}{$\chi$EFT\xspace}

\newcommand{\ptyp}{p_\text{typ}}
\newcommand{\LambdaEFT}{\overline{\Lambda}_\text{EFT}}
\newcommand{\LambdaNoPion}{\overline{\Lambda}_{\slashed{\pi}}}
\newcommand{\EFTNoPion}{EFT($\slashed{\pi}$)\xspace}
\newcommand{\LambdaChi}{\overline{\Lambda}_{\chi}}

\newcommand{\NXLO}[1]{N\ensuremath{{}^{#1}}LO\xspace}

\newcommand{\hq}{\hspace{0.5ex}}
\newcommand{\absatz}{\vspace{2ex}\noindent}

\newcommand{\myparagraph}[1]{\vspace*{1.1ex minus 0.5ex}\noindent\textbf{#1}\hspace*{2ex}}
\newcommand{\calA}{\mathcal{A}}
\newcommand{\calC}{\mathcal{C}}

\newcommand{\calO}{\mathcal{O}}

\newcommand{\wave}[3]{\ensuremath{{}^{#1}\mathrm{#2}_{#3}}}

\begin{document}

%%%%%%%%%%%%%%%%%%%%%%%%%%%%%%%%%%%%%%%%%%%%%%%%%%%%%%%%%%%%%%%%%%%
\section{Motivation: Serious Theorists Have Error Bars}

That our understanding of natural phenomena is based on concrete, falsifiable
predictions is deeply ingrained in the scientific method.
It is insufficient to compare numbers; one also must judge their reliability.
And since we do not trust experiments without error bars, why should it be
acceptable for a theorist to not assess uncertainties in a calculation
independently of the data to be explained?  Simply stating that this
is ``difficult'' is certainly no sufficient excuse, especially after the
recent surge of articles on theory errors; see e.g.~\cite{editorial,
  JPhysG}. The prospect of a reproducible, objective, quantitative estimate of
theoretical uncertainties lies thus at the heart not only of any Effective
Field Theory (EFT). But EFTs claim to possess well-defined schemes to find
just that. It is therefore befitting to explore how the validity of such
prescriptions can be gauged.

More than a decade ago, Lepage used a highly influential lecture to quantify
convergence to data~\cite{Lepage:1997cs}. In contradistinction, the test
presented here hopes to quantify the internal consistency of an EFT and takes
\emph{minimal resort} to experimental information. Ideally, theorists would
perform ``double-blind'' calculations in which theory errors are assessed
under the pretence that no or only very limited data is available. Such
``post-dictions'' are of course predictions when information is indeed
experimentally unknown or hard to access, or when data consistency must be
checked.

EFT uses scale-separation to expand interactions and observables in a
dimension-less quantity
\begin{equation}
  Q=\frac{\text{typical low momenta } k,\ptyp}{\text{breakdown scale } \LambdaEFT}<1\;\;.
\end{equation}
The numerator contains the relative momentum $k$ between scattering particles
and other intrinsic low scales, here summarily denoted by $\ptyp$. At the
breakdown scale $\LambdaEFT$, new dynamical degrees of freedom enter which are
not explicitly accounted but whose effects at these short distances are
simplified into Low-Energy Coefficients (LECs). In \ChiEFT, $\ptyp$ includes
the pion mass and the inverse scattering lengths of the $\mathrm{NN}$ system.
$\LambdaChi\approx[700\dots1000]\;\MeV$ is consistent with the masses of the
$\omega$ and $\rho$ as the next-lightest exchange mesons, and with the chiral
symmetry breaking scale.  The power counting (PC) is then determined by
Na\"ive Dimensional Analysis~\cite{NDA,NDA2}.  When all interactions are
perturbative, as in the mesonic and one-baryon sectors, this amounts to not
much more than counting powers of $k$ and $\ptyp$.

The situation is more complicated when some interactions must be treated
non-perturbatively at leading order because of shallow real or virtual bound
states with scales $\ptyp\ll\LambdaEFT$ in the EFT's range of validity. In
$\mathrm{NN}$ scattering, all terms in the LO Lippmann-Schwinger equation,
including the potential, must be of the same order when all nucleons are close
to their non-relativistic mass-shell. If not, one term could be treated as
perturbation of the others and there would be no shallow bound-state. Picking
the nucleon-pole in the energy-integration, $q_0\sim\frac{q^2}{2M}$, leads to
the consistency condition that both amplitude $T_\mathrm{NN}$ (ellipse) and
potential $V_\mathrm{NN}$ (rectangle) must be of order $Q^{-1}$:
\begin{equation}
  \label{eq:consistency}
  \begin{split}
    \includegraphics[clip=]{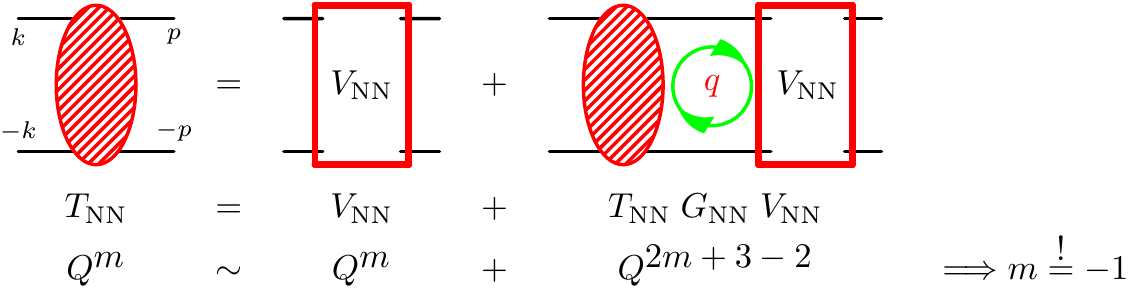}
  \end{split}
\end{equation}
In particular, the one-pion exchange appears to scale as
$(\vec{\sigma}_1\cdot\vec{q})(\vec{\sigma}_2\cdot\vec{q})/(\vec{q}^2+\mpi^2)\sim
Q^0$ when one counts only explicit low-momentum scales, but must be of order
$Q^{-1}$ if its iteration is to be mandated.
In a straightforward extension, amplitude and interaction between $n$
nucleons scale as
\begin{equation}
  \label{eq:consistency-n}
  T_{n\mathrm{N}}\sim V_{n\mathrm{N}}\sim Q^{1-n}
\end{equation} if it
is nonperturbative at LO.
Since both the interactions and the LECs themselves carry inverse powers of
$Q$, finding their importance by counting momenta is insufficient. This result
only assumes the existence of unnaturally small scales, irrespective of the
form of the interaction. It does not reveal \emph{which} terms constitute the
LO potential; only how those terms must be power-counted.

This behaviour has been long-recognised in pionless EFT (\EFTNoPion) and its
variants (like Halo-EFT and EFT of point-like interactions), where the scaling
of operators and the $\beta$ functions of couplings between up to 3 nucleons
are well-established~\cite{Bedaque:2002mn, Platter:2009gz}. For example,
analytic results in well-controlled limits show one momentum-independent
$3N$ operator at LO. Likewise, non-relativistic QED and QCD count the Coulomb
potential as $Q^{-1}$ to allow its resummation.

The situation in \ChiEFT for two and more nucleons is less obvious. Weinberg
suggested to still count LECs as $Q^0$, and to apply the perturbative counting
of momenta not to amplitudes but to the few-nucleon potential, which is then
iterated to produce shallow bound states.  How this translates into a PC of
observables is under dispute. Further disagreement persists about the
interpretation of approximate solutions (large off-shell momenta,
semi-classical limit, etc.), and about unrelated numerical problems (deeply
bound states etc.). In addition, a cutoff $\Lambda$ becomes numerically
necessary. It is conceptually quite different from the breakdown scale
$\LambdaEFT$, albeit the two symbols are similar. It cannot be much smaller
than the breakdown scale in order not to ``cut out'' physical, low-resolution
momenta in loops. But even how far $\Lambda$ should be varied is under
dispute: Is any value $\Lambda\gtrsim\LambdaEFT$ legitimate, including
$\Lambda\to\infty$; or should the range be constrained to
$\Lambda\approx\LambdaEFT$?

%%%%%%%%%%%%%%%%%%%%%%%%%%%%%%%%%
\begin{table}[!b]
  \vspace*{-0.1ex}
  \centering\tiny\newcommand{\makerow}[5]{#1&#2&#5&#4&#3} \begin{tabular}{|l||p{0.18\linewidth}|p{0.187\linewidth}|p{0.215\linewidth}|p{0.205\linewidth}|}
    \hline
    \makerow{\rule[-1.2ex]{0pt}{4ex}\textbf{order}}
    {\textbf{Weinberg} (modified)~\cite{Weinberg}}
    {\textbf{Long/Yang}~2012 \cite{Long:2011xw, Long:2012ve}}
    {\textbf{Pavon et al.}~2006 \cite{PavonValderrama:2005wv,
        PavonValderrama:2005uj, Valderrama:2009ei, Valderrama:2011mv}}
    {\textbf{Birse} 2005 \cite{Birse:2005um, Birse:2009my}}
    \\\hline\hline
    \makerow{\rule[-1.2ex]{0pt}{4ex}{$Q^{-1}$}}{ 
      LO of \wave{1}{S}{0}, \wave{3}{S}{1}, OPE}{LO of \wave{1}{S}{0}, \wave{3}{S}{1}, OPE, \wave{3}{P}{0,2}}{
      LO of \wave{1}{S}{0}, \wave{3}{S}{1}, OPE, \wave{3}{P}{0,2}, \wave{3}{D}{2}}{
      LO of \wave{1}{S}{0}, \wave{3}{S}{1}, OPE, \wave{3}{D}{1}, \wave{3}{SD}{1}
    }\\\hline
    \makerow{\rule[-1.2ex]{0pt}{5ex}$Q^{-\half}$}{none}{none}{
      LO of \wave{3}{SD}{1}, \wave{3}{D}{1}, \wave{3}{PF}{2}, \wave{3}{F}{2}}{
      LO of \wave{3}{P}{0,1,2}, \wave{3}{PF}{2}, \wave{3}{F}{2}, \wave{3}{D}{2}
    }\\\hline% \hline
    \makerow{\rule[-1.2ex]{0pt}{4ex}$Q^{0}$}{none}{
      NLO of \wave{1}{S}{0}}{NLO of \wave{1}{S}{0}}{NLO of \wave{1}{S}{0}
    }\\\hline
    \makerow{\rule[-1.2ex]{0pt}{5ex}$Q^{\half}$}{none }{none}{none}{ NLO of \wave{3}{S}{1}, \wave{3}{D}{1}, \wave{3}{SD}{1} }\\
    \hline
    \makerow{\rule[-1.2ex]{0pt}{4ex}\multirow{2}{*}{$Q^1$}}{LO of \wave{3}{SD}{1},\wave{1}{P}{1}, \wave{3}{P}{0,1,2},
        TPE;
      NLO of \wave{1}{S}{0}, \wave{3}{S}{1}}{LO of \wave{3}{SD}{1},\wave{1}{P}{1}, \wave{3}{P}{1},
      \wave{3}{PF}{2}, TPE;
      NLO of  \wave{3}{S}{1}, \wave{3}{P}{0}, \wave{3}{P}{2};
      \NXLO{2} of \wave{1}{S}{0}
    }{\multirow{2}{*}{none}}{\multirow{2}{*}{none}}\\\hline
    \makerow{\rule[-1.2ex]{0pt}{5ex}$Q^\frac{3}{2}$}{none}{none
    }{none}{NLO of \wave{3}{D}{2}, \wave{3}{P}{0,1,2}, \wave{3}{PF}{2}, \wave{3}{F}{2}
    }\\\hline
    \makerow{\rule[-1.2ex]{0pt}{4ex}\multirow{2}{*}{$Q^2$}}{\multirow{2}{*}{NLO of TPE}}{ \multirow{2}{*}{NLO of TPE; \NXLO{3} of \wave{1}{S}{0}}}{
      LO of TPE, \wave{1,3}{P}{1};
      NLO of \wave{3}{S}{1}, \wave{3}{D}{1,2}, \wave{3}{SD}{1}, \wave{3}{P}{0,2},
      \wave{3}{PF}{2}; \NXLO{2} of
      \wave{1}{S}{0}}{
      LO of TPE, \wave{1}{P}{1}; NLO of OPE;  \NXLO{2} of
      \wave{1}{S}{0}}\\\hline
    \hline\hline
    \makerow{\rule[-1.2ex]{0pt}{4ex}  {\#    at $Q^{-1}$}}{$2$}{$4$}{$5$}{$4$}\\\hline
    \makerow{\rule[-1.2ex]{0pt}{4ex} {\#     at $Q^{0}$}}{$+0$}{$+1$}{$+5$}{$+7$}\\\hline
    \makerow{\rule[-1.2ex]{0pt}{4ex}  {\#     at $Q^{1}$}}{$+7$}{$+8$}{0}{$+3$}\\\hline\hline
    \makerow{\rule[-1.2ex]{0pt}{4ex}total at $Q^{1}$}{$9$}{$13$}{$10$}{$14$}\\
    \hline
  \end{tabular}
  \caption{\label{tab:pc} Order $Q^n$ at which some LECs and the One- as well as
    Two-Pion-Exchange (OPE, TPE) enter in partial waves, for
    proposed power-countings in $\mathrm{NN}$  \ChiEFT~\cite{Saclay}. LECs
    of mixing angles are denoted e.g.~by \wave{3}{SD}{1}. The bottom part
    summarises the number of LECs  at a given order. Not all  schemes
    have contributions at an order, and some do not list all
    higher partial waves. While the information was collected with
    feedback from the respective authors, only I am to blame for
    errors. The results of Weinberg's PC have been shifted by $-1$ so that its
    potential  starts at order $Q^{-1}$, as mandated by the general arguments of
    eq.~\protect\eqref{eq:consistency}. %Fig.~\protect\ref{fig:consistency}.
  }
\end{table}
%%%%%%%%%%%%%%%%%%%%%%%%%%%%%%%%%
It is thus no surprise that four active PC proposals emerged in \ChiEFT, all
with the same degrees of freedoms: nucleons and pions only~\cite{Weinberg,
  Birse:2005um, Birse:2009my, PavonValderrama:2005wv, PavonValderrama:2005uj, Valderrama:2009ei, Valderrama:2011mv, Long:2011xw,
  Long:2012ve}. Table~\ref{tab:pc} lists their predictions for the order at
which a LEC enters in the lower $\mathrm{NN}$ partial waves. Each finds a
different number of LECs at any given order -- and each claims consistency.
Not all can be right, though.
Articles, panels and sessions at Chiral Dynamics and other conferences as well
as dedicated workshops led to little consensus (see e.g.~Daniel Phillips'
even-handed account at the last Chiral Dynamics~\cite{Phillips:2013fia}); some
additional notable contributions include Refs.~\cite{Kaplan:1996xu,
  Kaplan:1998we, Beane:2001bc, Nogga:2005hy, Epelbaum:2006pt, Beane:2008bt, 
  Epelbaum:2009sd}.

This is not just stamp-collecting or a philosophical question which
potentially exposes the soft underbelly of \ChiEFT and the credibility of its
error assessments, but which is ``otherwise'' of little practical
consequence. A central EFT promise is that it encodes the unresolved
short-distance information at a given accuracy into not just some, but the
\emph{smallest-possible} number of independent LECs. For example, the PC
proposals of $\mathrm{NN}$ \ChiEFT differ most for attractive triplet waves:
the \wave{3}{P}{2}-\wave{3}{F}{2} system at order $Q^0$ has no LEC
parameter~\cite{Weinberg} -- or $3$ of similar size~\cite{Birse:2005um,
  Birse:2009my} -- or $3$, but with different
weights~\cite{PavonValderrama:2005wv, PavonValderrama:2005uj,
  Valderrama:2009ei, Valderrama:2011mv} -- or $1$~\cite{Long:2011xw,
  Long:2012ve}. To bring it to a boil: If all proposals are renormalised and
fit $\mathrm{NN}$ data with the same $\chi^2$, the one with the least number
of parameters wins.

For the sake of this note, I am agnostic about this dispute. Rather, I propose
to test if a predicted convergence pattern is reflected in the answers,
i.e.~if a proposed power counting is consistent. % As alluded to at
% the beginning, a consistent power-counting is central to any EFT: only when
% all contributions can be ordered by relative size before calculating
% specific terms, can residual uncertainties be estimated reliably. So, how to
% guarantee that the assumptions of a power counting bear out in the final
% results?

\absatz On a historical note, the origin of these remarks goes back to
publications in 2003 and 2005~\cite{Bedaque:2002yg, improve3body}, and to
lectures at the 2008 US National Nuclear Physics Summer
School~\cite{NNPSS}. When the issue was revisited at two more recent
workshops~\cite{Saclay, Benasque}, its conclusions were generally perceived as
not immediately straightforward or widely known. Input on some aspects was
also provided for two recent publications~\cite{Furnstahl:2014xsa,
  Epelbaum:2014efa}. It seems therefore fitting to present the test in the
form of an expanded Technical Note. This article summarises an upcoming
publication~\cite{hgrie}.

%%%%%%%%%%%%%%%%%%%%%%%%%%%%%%%
\section{The Test: Turning Cutoff Dependence into an Advantage}

Assume we calculated an observable $\calO$ up to and including order $Q^n$ in
an EFT:
\begin{equation}
  \label{eq:observable}
  \calO(k,\ptyp;\Lambda;\LambdaEFT)=
  \sum\limits_i^n\left(\frac{k,\ptyp}{\LambdaEFT}\right)^i
  \calO_i(k,\ptyp;\LambdaEFT)\; +\;\calC_n(\Lambda;k,\ptyp,\LambdaEFT)\left(\frac{k,\ptyp}{\LambdaEFT}\right)^{n+1}
\end{equation}
[Non-integer $n$ and non-integer steps from one order to the next will be discussed in Sect.~\ref{sec:principle}.]
The notation indicates that numerators may depend on both $k$ and $\ptyp$. In
a properly renormalised result, effects attributed to a regulator $\Lambda$
appear only at orders which are higher than the last order $n$ which is known
in full.
% and grow at most like $(k,\ptyp)/\Lambda$, adding one more power of $k$.
% Notice that no particular assumption is necessary as to the size of
% $\Lambda$ relative to $\LambdaEFT$. The theory is however of course only
% sensible if all loop momenta are sampled which lie in the domain of validity
% of the EFT, i.e.~if $\Lambda\gtrsim\LambdaEFT$. Only then can the
% coefficients $\calC_n$ be expected to be of natural size relative to
% $\LambdaEFT$ (with the caveats mentioned around
% eq.~\eqref{eq:consistency-n}).
%
The residual $\calC_n(\Lambda;k,\ptyp,\LambdaEFT)$ may still depend on
$\LambdaEFT$, $k$ and $\ptyp$, but it should be of natural size for all
$k,\ptyp<\LambdaEFT$, so that its contribution is parametrically suppressed by
$\left(\frac{k,\ptyp}{\LambdaEFT}\right)^{n+1}$ relative to the known terms of
the series. If not, cutoff variations produce corrections which are comparable
in size to the regulator-independent terms $\calO_i(k,\ptyp;\LambdaEFT)$ and
contradict the EFT assumption that higher-order corrections are parametrically
small.

The relative difference of $\calO(k,\ptyp;\Lambda)$ at any two cutoffs is
then:
\begin{equation}
  \label{eq:master}
  \frac{\calO_n(k,\ptyp;\Lambda_1)-\calO_n(k,\ptyp;\Lambda_2)}
  {\calO_n(k,\ptyp;\Lambda_1)}=\left(\frac{k,\ptyp}{\LambdaEFT}\right)^{n+1}
  \times\;\frac{\calC_n(\Lambda_1;k,\ptyp,\LambdaEFT)-
    \calC_n(\Lambda_2;k,\ptyp,\LambdaEFT)}
  {\calC_n(\Lambda_1;k,\ptyp,\LambdaEFT)}\;\;.
\end{equation}
One can now vary $k$ or $\ptyp$ to read off both the order $n$ to which the
calculation is complete and the breakdown scale $\LambdaEFT$ -- if the cutoff
behaviour cannot be eliminated in its entirety,
i.e.~$\calC_n(\Lambda_1)\ne\calC_n(\Lambda_2)$ for at least some cutoff pairs,
and if the residuals $\calC_n$ vary more slowly with $k$ and $\ptyp$ than with
$\Lambda$.
The results of such a fit may certainly be inconclusive; see extended remarks
in Sect.~\ref{sec:notesofnote}.
But if higher orders are \emph{not} parametrically suppressed and the slope
comes out \emph{smaller} than the PC prediction $n+1$, then the EFT is
necessarily not properly renormalised. As will be discussed in
Sect.~\ref{sec:principle}, a slope $\ge n+1$ does not suffice to demonstrate
consistency or establish failure.

Equation \eqref{eq:master} is formulated in terms of renormalised quantities
only and therefore holds for any regulator, but it is most useful for cutoffs:
Answers in nonperturbative EFTs are usually found only numerically and for a
\emph{cutoff regulator}, i.e.~for a regulator which explicitly suppresses
high momenta $q\gtrsim\Lambda$ in loops.  It is this case which we use to our
advantage from now on.

Cutoffs are of course only sensible if all loop momenta are sampled which lie
in the domain of validity of the EFT, i.e.~if $\Lambda\gtrsim\LambdaEFT$.
Only then can the coefficients $\calC_n$ be expected to be of natural size
relative to $\LambdaEFT$ (with the caveats mentioned around
eq.~\eqref{eq:consistency-n}). Except for this, no particular assumption is
necessary as to the size of $\Lambda$ relative to $\LambdaEFT$ in
eq.~\eqref{eq:master}.  In dimensional regularisation and other analytic
schemes, on the other hand, renormalisation can be performed exactly and no
cutoff $\Lambda$ or residual regulator scale appears in observables at all.
Equation \eqref{eq:master} is then an exact zero, with no information about
$n$ and $\LambdaEFT$. But doubts about proper renormalisation of a calculation
which is analytic at each step do not arise, so the test is moot anyway.

%%%%%%%%%%%%%%%%%%%%%%%%%%%%%%%
\section{An Application: Confirming the Hierarchy of $3N$ Interactions in \EFTNoPion}
\label{sec:application}

Before continuing the discussion of the parameters of the test in
Sect.~\ref{sec:notesofnote}, consider the first application (to my knowledge)
of this test: the \wave{2}{S}{\frac{1}{2}} $Nd$ wave in \EFTNoPion. It is
well-known that its $3N$ interaction without derivatives does not follow
simplistic PC rules (``just count momenta'') which predict $H_0$ at \NXLO{2}
or $\calO(Q^0)$~\cite{Bedaque:2002mn, Platter:2009gz}. Instead, it is needed
at LO to stabilise the system (Thomas-collapse, Efimov effect); its scaling,
$H_0\sim Q^{-2}$, follows from eq.~\eqref{eq:consistency-n} for $n=3$. If the
first momentum-dependent $3N$ interaction $k^2H_2$ follows the simplistic
argument and scales as $Q^2$, then new LECs need to be determined from $3N$
data only at \NXLO{4}. Therefore, one could find $2N$ interaction strengths
from few-N data with only one new $3N$ datum up to an accuracy of better than
$1$\% at low momenta.  This is crucial for example for hadronic
flavour-conserving parity violation since it considerably extends the number
of targets and observables~\cite{Griesshammer:2010nd}.

Based on the asymptotic off-shell amplitudes, Refs.~\cite{Bedaque:2002yg,
  improve3body} proposed that $H_2$ is only suppressed by $Q^2$ relative to
LO, i.e.~that calculations at \NXLO{2} or on the $10$\%-level do already need
one additional $3N$ datum as input. In Ref.~\cite{effrange}, this was
confirmed and extended to a general scheme to find the order at which any
given $3N$ interaction starts contributing. The argument analyses
perturbations to the asymptotic form of the LO integral equation. It
%procedure 
is not immediately transparent, as witnessed by a subsequent claim that a
$k$-dependent $3N$ interaction enters not earlier than
\NXLO{3}~\cite{Platter:2006ev}. Upon closer inspection, it was later
refuted~\cite{Ji:2012nj}.

\begin{figure}[!h]
  \centering \parbox{0.59\linewidth}{\includegraphics[width=\linewidth]%[height=0.315\linewidth]%[width=0.7\linewidth]
    {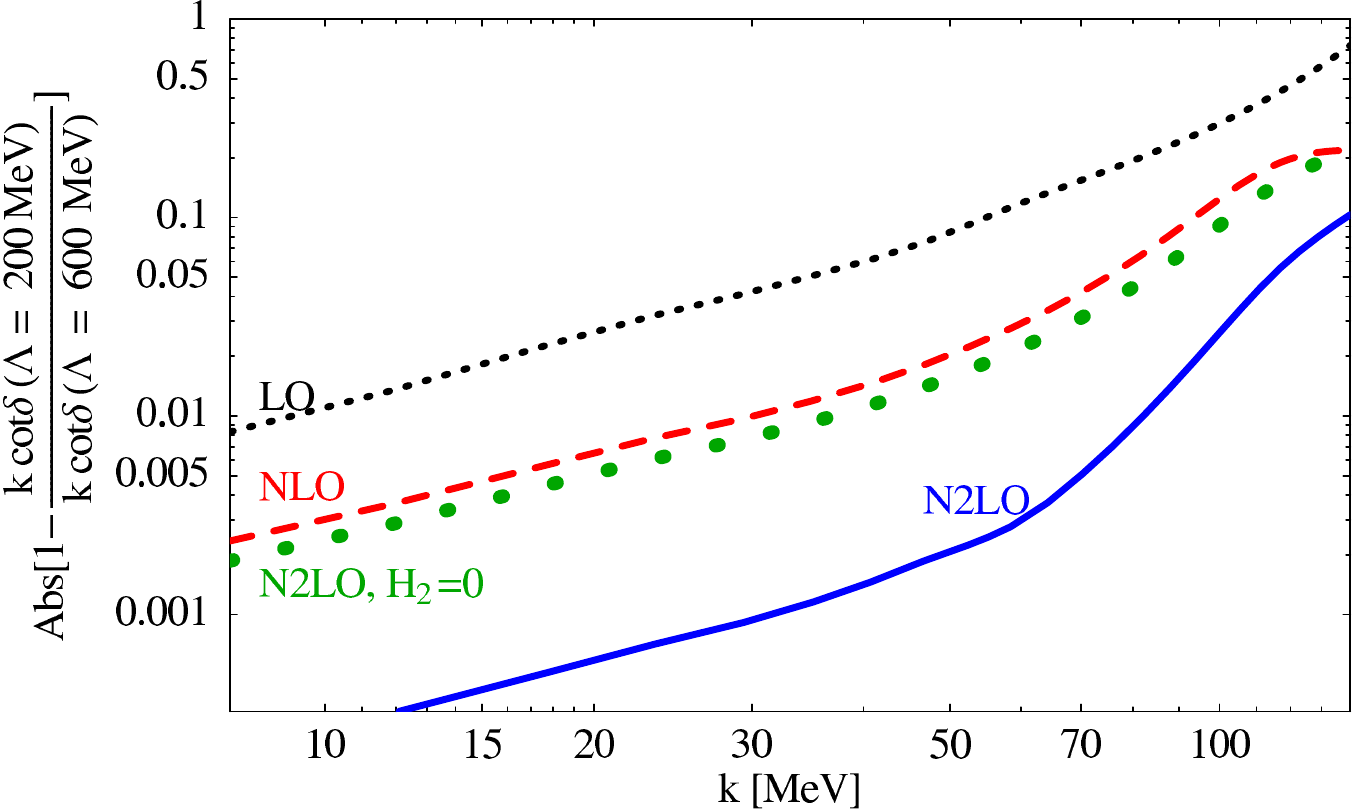}}
  \hspace*{2ex} \parbox{0.375\linewidth}{\includegraphics[width=\linewidth]%[height=0.3\linewidth]
    {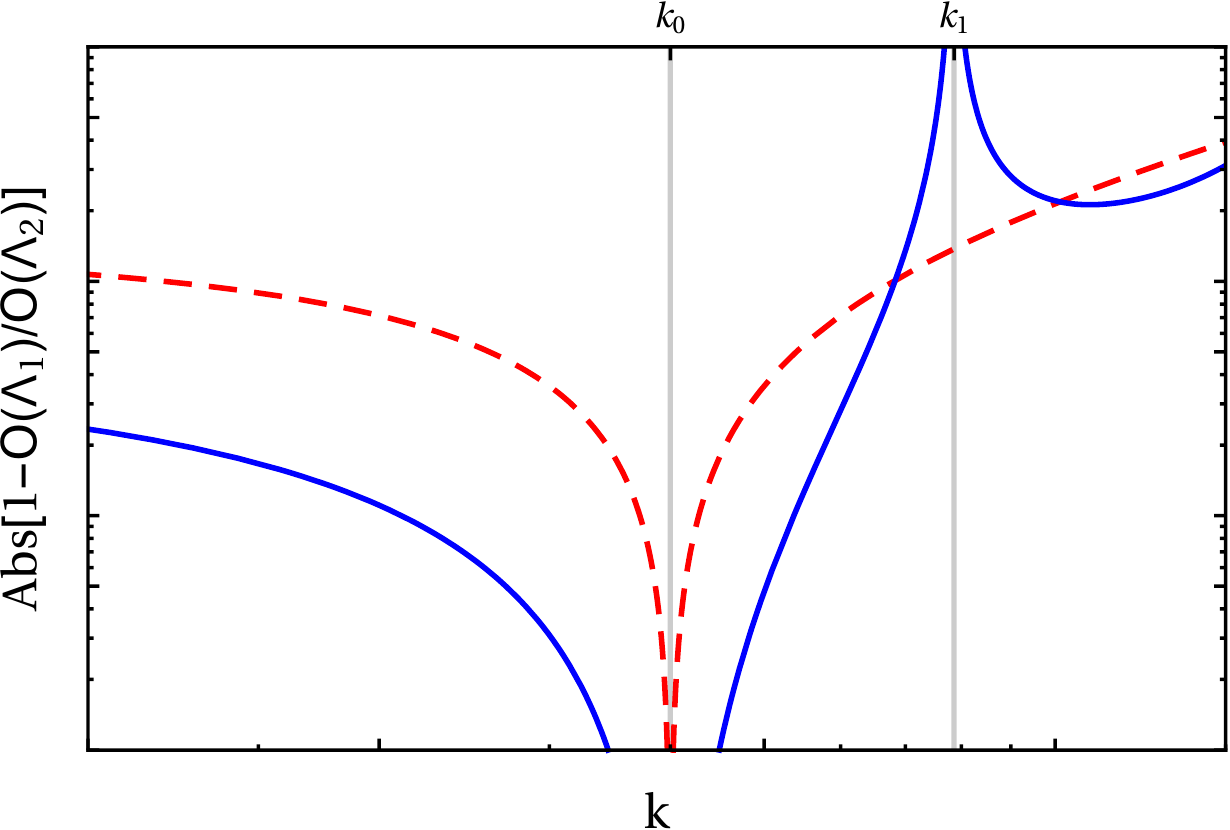}}
  \caption{\label{fig:nonLepage} \emph{Left}: Double-logarithmic error plot
    for the \wave{2}{S}{\frac{1}{2}} wave of $Nd$ scattering in \EFTNoPion;
    cf.~Refs.~\cite{Bedaque:2002yg, improve3body}. \label{fig:qualitative}
    \emph{Right}: Qualitative example of the impact of zeroes in
    $\calO(\Lambda_2)-\calO(\Lambda_1)$ (exact reproduction of datum at
    $k_0$), and in $\calO(\Lambda)$ (``accidental zero'' of $\calO(\Lambda_2)$
    at $k_1$).  Red dashed line: $n=1$; blue solid: $n=2$.}
\end{figure}

Refs.~\cite{Bedaque:2002yg, improve3body} also supplied numerical evidence
from solutions of the Faddeev equations in momentum space with a step-function
cutoff: a double-logarithmic plot of eq.~\eqref{eq:master} for the inverse $K$
matrix, $\calO=k\cot\delta$ at $\Lambda_1=900\;\MeV$ and
$\Lambda_2=200\;\MeV$, both well above the breakdown scale
$\LambdaNoPion\approx\mpi$ of \EFTNoPion. A slight variant is reproduced here
as Fig.~\ref{fig:nonLepage}. The cutoff dependence decreases order-by-order as
expected when the theory is perturbatively renormalised in the EFT
sense. There is no decrease from NLO to \NXLO{2} when $H_2\equiv0$. That by
itself could be accidental -- after all, would one not expect better
convergence with one more parameter to tune?

More informative is a look at the slopes. Lines at different orders are
near-parallel for small $k$ because there are additional natural low-energy
scales $\ptyp$, namely the binding momenta of the deuteron
($\gamma_t\approx45\;\MeV$) and of the virtual singlet-S state
($\gamma_s\approx8\;\MeV$). For $k\lesssim\gamma_{t,s}$, eq.~\eqref{eq:master}
is not very sensitive to $k$, so all slopes should be small and
near-identical. However, when $k\gg\gamma_{t,s}$ (but of course still
$k\ll\LambdaNoPion$, so that the EFT converges), they converge towards one
region.

Indeed, the fits of $n$ to the nearly straight lines in the momentum range
between $70$ and $100$ to $130\;\MeV<\LambdaNoPion$ compare well to the PC
prediction when $H_2$ is added at \NXLO{2}~\cite{improve3body}:
\vspace{-0.5ex}
\begin{equation}
\label{eq:3Ntab}
  \begin{tabular}{c@{\hq}|@{\hq\hq}c@{\hq\hq}c@{\hq\hq}c@{\hq\hq}|@{\hq\hq}c}
    & LO&NLO&\NXLO{2}&\NXLO{2} {without $H_2$}\\%[1ex]
    \hline
    {$n+1$} fitted &${\sim 1.9}$&{$2.9$}&{$4.8$} [\emph{sic!}]&${3.1}$\\%[1ex]
    {$n+1$} predicted &${2}$&{$3$}&{$4$}&not renormalised
  \end{tabular} 
\end{equation}
Without $H_2$ at \NXLO{2}, the slope does not improve from NLO. This is a
clear signal that the PC is inconsistent without a momentum-dependent $3N$
interaction at \NXLO{2}: Its assumptions do not bear out in the functional
behaviour of this observable on $k$.  On the other hand, when $H_2$ is
included, the slope is markedly steeper than at NLO. The general agreement
between predicted and fitted slope is astounding, and actually quite stable
against variation of the fit range or of the two cutoffs $\Lambda_1$ and
$\Lambda_2$. Only the LO numbers are somewhat sensitive, and only to the upper
limit~\cite{improve3body}.

It is somewhat surprising that the slope increases by two units from NLO to
\NXLO{2} when one includes $H_2$. One would have expected the change from each
order to the next to be by only one unit. This may stem from the ``partially
resummed formalism'' used at that time. Since that resums some higher-order
contributions, it may be worth revisiting this issue with J.~Vanasse's method
to determine higher-order corrections in ``strict
perturbation''~\cite{Vanasse:2013sda}. But we will see in the notes on
``Assumptions of the Expansion'' in Sect.~\ref{sec:notesofnote} that a fitted
slope which is larger than predicted does not invalidate the power counting --
the converse does.

Finally, one reads off a rough value of
$\LambdaNoPion\approx[120\dots150]\;\MeV$ as the region where the fitted lines
coalesce. This is not in disagreement with the breakdown scale expected of
\EFTNoPion.

%%%%%%%%%%%%%%%%%%%%%%%%%%%%%%%
\section{Notes of Note}
\label{sec:notesofnote}

With this example in mind, let us consider assumptions, strengths, extensions,
features, caveats and limitations of such an analysis to assess the
consistency of a PC proposal.

\subsection{Matters of Principle}
\label{sec:principle}

\myparagraph{Renormalisation Group Evolution} Multiply eq.~\eqref{eq:master}
by $(\Lambda_1-\Lambda_2)/\Lambda_1$ and take $\Lambda_2\to\Lambda_1$:
\begin{equation}
  \label{eq:rge} \frac{\Lambda}{\calO}\;\frac{\dd\calO}{\dd\Lambda}=
  \left(\frac{k,\ptyp}{\LambdaEFT}\right)^{n+1}
  \;\frac{\dd\ln\calC_n(\Lambda)}{\dd\ln\Lambda}\;\;.
\end{equation}
This is Wilson's Renormalisation Group Equation for the observable $\calO$.
Note that eq.~\eqref{eq:rge} features a \emph{total} derivative: LECs in
$\calO$ are readjusted as $\Lambda$ changes. 

In practise, an EFT at finite order $n$ and with finite cutoff tolerates
cutoff artefacts which are parametrically small, i.e.~at least of order
$n+1$. This also limits the rate of change in the residual $\calC_n$: I call
an observable ``perturbatively renormalised'' when the right-hand side of
eq.~\eqref{eq:master} is smaller than any term on the left-hand side. To some,
this condition implies $\Lambda$ can only be varied in a range around
$\LambdaEFT$; the functional dependence on $k$ and $n$ is then still a
quantitative prediction. A double-logarithmic plot reveals quantitative
aspects of the Renormalisation Group evolution and can be utilised to falsify
claims of consistency in an EFT.

\myparagraph{Extending the Expansion} The order $n$ is not counted relative to
LO. It is not even necessarily an integer, as Table~\ref{tab:pc} shows, and
the first omitted order is not always $Q^{n+1}$, but more generally
$Q^{n+\alpha}$, $\mathrm{Re}[\alpha]>0$. To replace $n+1\to n+\alpha$ in
eqs.~\eqref{eq:observable}, \eqref{eq:master}, \eqref{eq:rge} -- and indeed
throughout -- is straightforward. In \EFTNoPion, the slope-fit in
eq.~\eqref{eq:3Ntab} endorses that the $3N$ PC proceeds in integer steps.
Including non-analytic dependencies of the residuals on $k$ or $\ptyp$ is also
straightforward. For the remainder of the presentation, all such replacements
are implied, but we stick to the integer case for convenience.

\myparagraph{Assumptions of the Expansion} The assumptions on the residual
$\calC_n$ are endorsed if order $n$ and breakdown scale $\LambdaEFT$ follow
indeed the functional form of eq.~\eqref{eq:master} or its
variant~\eqref{eq:rge}. Na\"ive Dimensional Analysis (NDA) sets the magnitude
of $\calC_n$ to the scale of its running~\cite{NDA, NDA2}. Its
cutoff-dependence and other effects are eventually absorbed into higher-order
LECs %, i.e.~the cutoff dependence of observables should generically decrease
% order-by-order -- even when no new fit parameters/LECs are encountered
(see also below).

We can actually be somewhat more specific about the condition that the
variation of the residual $\calC_n$ variation with respect to $\Lambda$ should
be larger than that for other parameters. Since
$k,\ptyp\ll\Lambda_1,\Lambda_2$, the dimensionless ratio on the left-hand side
of eq.~\eqref{eq:master} can be expanded as
\begin{equation} 
\label{eq:residualexpansion}
  \frac{\calC_n(\Lambda_1;k,\ptyp,\LambdaEFT)-
    \calC_n(\Lambda_2;k,\ptyp,\LambdaEFT)} 
  {\calC_n(\Lambda_1;k,\ptyp,\LambdaEFT)}= c_0(\Lambda_1,\Lambda_2;\LambdaEFT)+
  c_1(\Lambda_1,\Lambda_2;\LambdaEFT)\;\frac{k,\ptyp}{\Lambda_1,\Lambda_2}+\dots
\end{equation}
If the first term dominates, then the dependence of eq.~\eqref{eq:master} on
$k$ and $\ptyp$ is indeed indicative of the order $Q^{n+1}$. If subsequent
terms dominate, the slope may be larger than $n+1$ -- but never smaller. 

\myparagraph{Necessary but Not Sufficient} This shows that a slope smaller
than $n+1$ conclusively demonstrates failure of the PC to be
consistent. However, the criterion is necessary rather than sufficient:
Slopes $\ge n+1$ are proof neither of failure, nor of success. Indeed, a PC may
be inconsistent but the coefficient of the terms with slope $<n+1$ may be
anomalously small, leading to a ``false positive''. 

\myparagraph{Estimating the Expansion Parameter} When the cutoffs $\Lambda_1$
and $\Lambda_2$ are both varied over a wide range\footnote{Some claim that
  renormalisability requires that $\calO$ has a unique limit as
  $\Lambda\to\infty$.}, the analysis also gives a practical way to find the
size of the expansion parameter as a function of $k$. Ratios between different
orders estimate $Q(k,\ptyp)$, and hence residual theoretical uncertainties as
function of $k$. This is of course only one way to assess $Q(k)$; within
reason, the least optimistic and hence most conservative of several methods
should be picked. For example, Ref.~\cite{Griesshammer:2011md} combined this
with the convergence pattern of the EFT series; see
also~\cite{Furnstahl:2015rha}.

\myparagraph{Choice of Expansion Parameter} In Sect.~\ref{sec:application},
$k$ is varied while the other scales $\ptyp$ are fixed, but any combination of
the low-energy scales may serve as variable(s). For example, scanning in the
pion mass at fixed $k\ll m_\pi$ may elucidate the $m_\pi$-dependence of some
couplings, with particular relevance to extrapolating lattice computations at
non-physical pion masses. Here, I will continue to concentrate on variations
with $k$, but most issues transfer straightforwardly to other variations.

\myparagraph{Window of Opportunity} One can read off slopes most easily in the
range $\ptyp<k<\LambdaEFT$. In \EFTNoPion, that window is narrow but suffices:
$\LambdaNoPion/(\ptyp\sim\gamma_{t,s})\lesssim3$. In \ChiEFT with dynamical
$\Delta(1232)$ degrees of freedom, we expect a wider range:
$\LambdaChi/(\ptyp\sim m_\pi)\gtrsim4$.  One may of course also fit the
variables $n$ and $\LambdaEFT$ in eq.~\eqref{eq:master} to the numerical
results outside that window, but then one needs to specify $\ptyp$ and
determine its contribution relative to $k$.

\myparagraph{Choice of Regulator} Residual cutoff dependence comes naturally
in numerical computations. This tests uses it as a tool to check
consistency. The example used a ``hard'' cutoff, but $\LambdaEFT$ and $n$ do
not depend on a specific regulator. If the theory can be renormalised exactly,
all residual regulator dependence disappears by dimensional transmutation;
cf.~\eqref{eq:rge}.

\begin{figure}[!htb]
  \centering
  \includegraphics[height=0.285\linewidth]{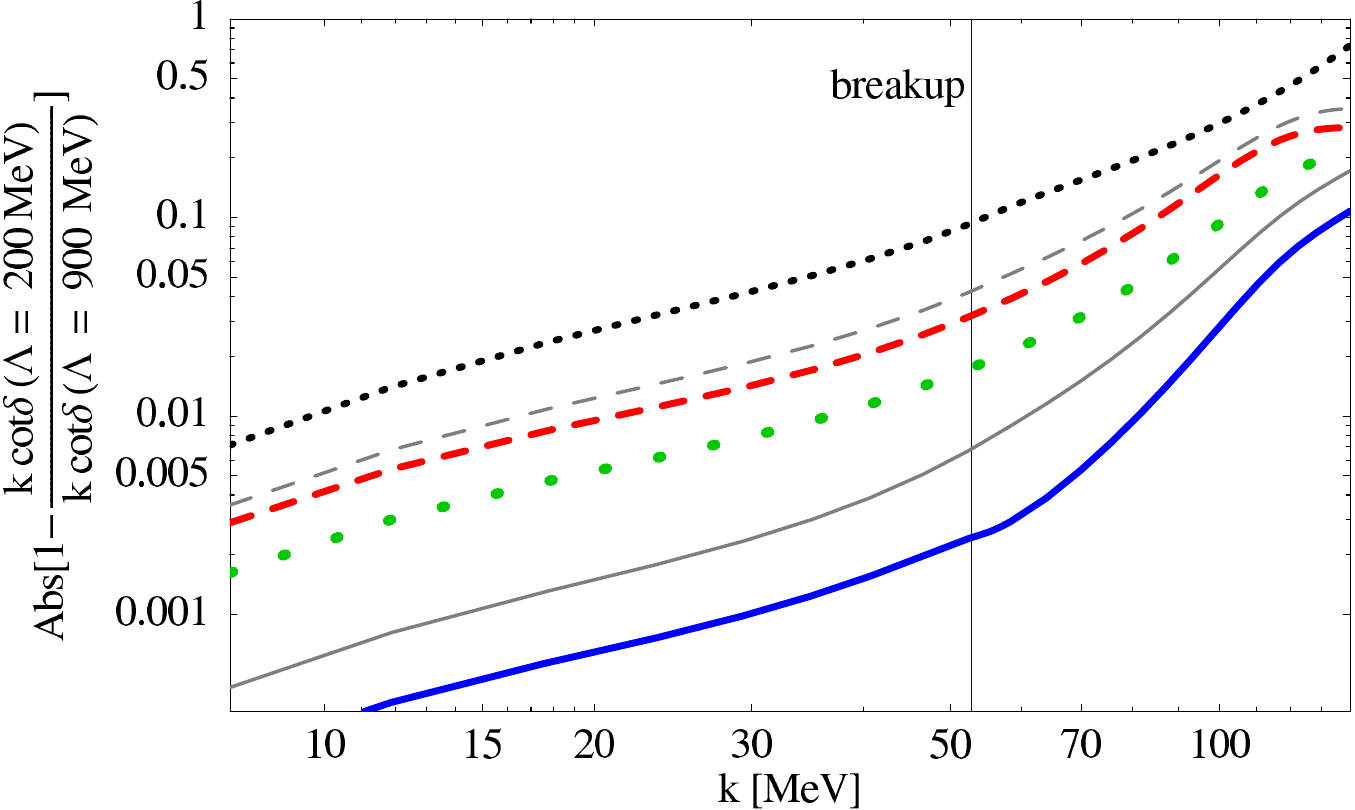}
  \hspace*{2ex}
  \includegraphics[height=0.285\linewidth]{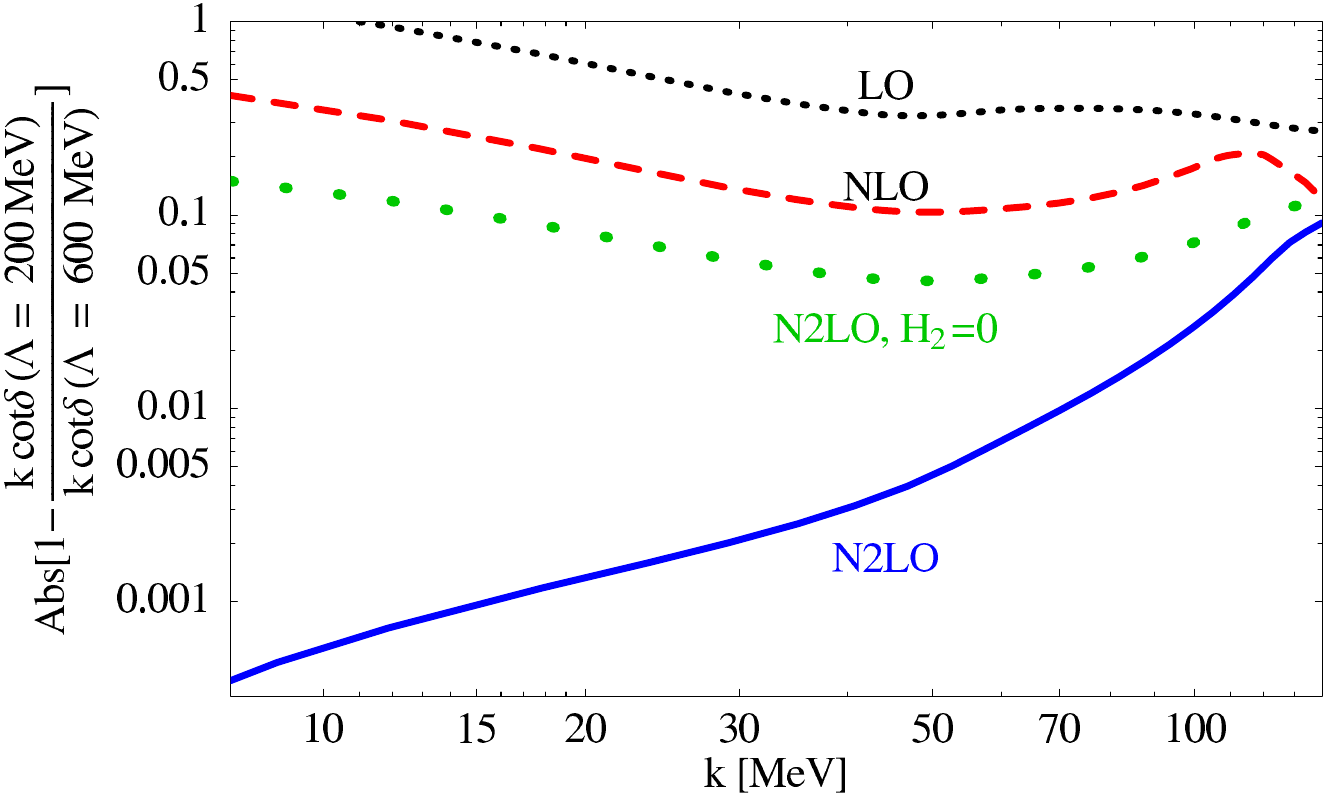}
  \caption{\label{fig:nonLepage2} \label{fig:nonLepage900MeV} \label{fig:EREvsZ}
    \emph{Left}: Thick coloured lines: Z-parametrisation of the $\mathrm{NN}$
    amplitude as in Fig.~\protect\ref{fig:nonLepage}, but for
    $\Lambda_1=900\;\MeV$, not $600\;\MeV$; thin gray lines: Bethe's Effective
    Range Parametrisation; from
    Ref.~\protect\cite{improve3body}. \label{fig:toB3} \emph{Right}: Test when
    the leading $3N$ interaction is determined not by the $Nd$ scattering
    length as in Fig.~\protect\ref{fig:nonLepage}, but by the position of the
    triton pole. The \NXLO{2} fit uses again the $Nd$ scattering length and
    triton binding energy.}
\end{figure}

\myparagraph{Choice of Cutoffs} The functional dependencies of
eqs.~\eqref{eq:master} and \eqref{eq:rge} on $n$ and $\LambdaEFT$ do not
depend on $\Lambda_1$ and $\Lambda_2$. While any two cutoffs
$\Lambda_1,\Lambda_2\gtrsim\LambdaEFT$ will do in principle, small leverage
may lead to numerical artefacts. The larger $\Lambda_2-\Lambda_1$, the clearer
the signal should be. For our example, Fig.~\ref{fig:nonLepage900MeV} shows
that an upper cutoff of $900\;\MeV$ instead of $600\;\MeV$ leads to different
curves but very similar slopes. Infinities, zeroes and oscillations of $\calO$
with $k$ for any pair $\Lambda_1,\Lambda_2$ can lead to problems (see
``\emph{Observables: Accidental Zeroes and Infinities}'' below) which are
readily avoided by choosing a cutoff pair such that
$\calO(\Lambda_1)-\calO(\Lambda_2)>0$ for all $k$. Even when one does not
choose to take one of the cutoffs to infinity\footnote{One could adhere to the
  philosophy that cutoffs and breakdown scales should be similar.}, a
reasonable range of allowed cutoffs exists. If $\Lambda_1\approx\Lambda_2$,
one may of course directly consider the numerical derivative of
eq.~\eqref{eq:rge} -- over a range of cutoffs. [To reiterate: exact cutoff
independence $\calO(\Lambda_1)\equiv\calO(\Lambda_2)$ for any cutoff pair is
not considered.]

\myparagraph{Decreasing Cutoff Dependence} Equation~\eqref{eq:master} is a
variant of the Renormalisation Group evolution of $\calO$, eq.~\eqref{eq:rge},
which in turn quantifies the fundamental EFT tenet that observables must
become order-by-order less sensitive to loop contributions beyond
$\LambdaEFT$, the range of applicability. Cutoff dependence in observables
should therefore generically decrease from order to order, irrespective
whether or not LECs are fitted. This does not apply to the $\calC_n$
themselves, but to the entire left-hand side of
eq.~\eqref{eq:master}. Refitting LECs may of course help to absorb some cutoff
dependence.
Indeed, no new LECs enter at NLO in the example above ($H_0$ is just
refitted), and the cutoff dependence decreases from LO to NLO. While it is
conceivable that the residual $\calC_n$ is sometimes somewhat larger than NDA
predicts, NDA should apply ``most of the time'', statistically speaking.

Still, a specific regulator form may produce a very small residual cutoff
dependence at one order but a significantly larger one at a subsequent order,
$\calC_n(\Lambda_1)-\calC_n(\Lambda_2)<
\calC_{n+1}(\Lambda_1)-\calC_{n+1}(\Lambda_2)$. This may for example occur if
the regulator produces only corrections with even powers of $\Lambda$ and the
numerics preserves this symmetry at least approximately (e.g.~because
$\Lambda_1\approx\Lambda_2$, allowing for a perturbative expansion). If this
overwhelms the expansion in $Q$, $\calO$ may indeed systematically become more
dependent on $\Lambda$ between some orders, but not between all. Nonetheless,
one should not just see some qualitatively improved cutoff dependence with
increasing order, but one must see the quantitatively predicted slopes emerge
for many orders: they must be $\ge n+1$; see eq.~\eqref{eq:residualexpansion}.

\myparagraph{Constructing a PC by Trial-and-Error} If the cutoff dependence of
a given observable does not decrease consistently between subsequent orders,
caution may be advisable. For example, $\Lambda$-dependence may increase from
one order to the next, but then decrease markedly when another full order with
a new LEC is included. This could signal that this LEC cures cutoff dependence
already at a lower order -- and hence that the PC is inconsistent. One should
then study the convergence pattern as the LEC is promoted to a lower order
such that the cutoff dependence decreases always between subsequent
orders. This may help to construct a consistent PC by trial-and-error and
iteration. Remember also that after a LEC starts contributing at a certain
order, it is re-adjusted at each subsequent order to absorb both cutoff
effects and still match its determining datum.

\myparagraph{Calculating Higher Orders} Traditionally, observables beyond LO
have been found by ``partially resumming'' contributions, i.e.~the
power-counted potential is iterated like in Weinberg's original
suggestion. Since corrections to the LO potential are defined as
parametrically small, they can be included in ``strict perturbation'',
avoiding potential problems with spurious deeply bound state which can be
generated by iteration~\cite{Vanasse:2013sda}. This may also provide clearer
signals for the PC test.

%%%%%%%%%%%%%%%%%%%%%%
\subsection{Picking Observables}

\myparagraph{Isolating Dynamical Effects} While any observable
could be chosen, those which are free from kinematic or other constraints
(e.g.~from symmetries) are preferred.
Consider the scattering amplitude $\calA_l$ in the $l$th partial wave (for
simplicity, assume no mixing). Since it is complex, one could choose
$\calO=|\calA_l|$. However, unitarity relates $\calA_l=1/(k\cot\delta_l-\ii
k)$ to the phase shift $\delta_l$. This constraint dominates when $\delta_l$
is between about $\pi/4$ and $3\pi/4$ -- which affects much of the
$\mathrm{NN}$ S-wave phase shifts. Even outside this interval, the additional
contribution to eq.~\eqref{eq:master} is not sensitive to dynamics.
In addition, analyticity dictates that phase shifts approach zero like
$k^{2l+1}$ for $k\to0$ in the $l$th partial wave. Since both numerator and
denominator in eq.~\eqref{eq:master} are then zero, $\calO=\delta_l$ is
dominated by numerical uncertainties as $k\to0$. This may not be a problem if
the region in which the slopes are determined is far away, but only a closer
inspection could tell if that holds.  Likewise, one eliminates phase-space
factors in decay constants, production cross sections, etc.

A sensible choice for single-channel scattering appears thus to be
$\calO=k^{2l+1}\cot\delta_l$: It is only constrained to be real below the
first inelasticity, and imaginary parts are usually small above it. Indeed,
the S-wave example above kept track of the imaginary part by plotting
\begin{equation}
  \left|1-\frac{k\cot\delta_0(\Lambda_2)}{k\cot\delta_0(\Lambda_1)}\right|\;\;.
\end{equation}
While factors of $k$ formally cancel, one computes $\calA$ ($k\cot\delta_0$),
so that numerics is more benign.

\myparagraph{Partial-Wave Mixing} In the $\mathrm{NN}$ system,
two partial waves with total angular momentum $J$ mix. The corresponding
unconstrained observables in the Stapp-Ypsilanti-Metropolis (SYM or
``nuclear-bar'') parametrisation are
\begin{equation}
  k^{2\pm1-2J}\bar{\delta}_{J\pm1}\;\;\mbox{ and }
  \;\;k^{-(2J+1)}\bar{\epsilon}_J\;\;.
\end{equation}
In the Blatt-Biedenharn parametrisation, the same rules apply for the
eigenphases, but $k^{-2}\epsilon_J$ is the unconstrained variable for the
mixing angle; see e.g.~\cite{deSwart:1995ui}. These choices do not suffer from
unitarity constraints (except for being real below the first inelasticity) and
can be used directly.

\myparagraph{Dependence on Parameter Input} Let us first consider
processes in which $\calO(k)$ is a parameter-free prediction, i.e.~its LECs
are all known from some other process(es). To what extent does the procedure
depend on that choice? In the example, the two-nucleon interactions were
determined to match the Z-parametrisation of $\mathrm{NN}$-scattering (fit to
pole position and residue of the scattering
amplitude)~\cite{Phillips:1999hh}. Fig.~\ref{fig:EREvsZ} shows that results
with Bethe's Effective-Range parametrisation have a markedly different rate of
convergence, but the extracted slopes and $\LambdaNoPion$ agree very
well~\cite{improve3body}.

\myparagraph{Accidental Zeroes and Infinities} Some observables
may show additional structures which should be avoided. For example, the
\wave{3}{P}{0} phase shift in $\mathrm{NN}$ scattering is zero at a lab energy
of about $150\;\MeV$, so that the relative deviation of $\calO=\delta_l$ in
eq.~\eqref{eq:master} diverges. Likewise, $\calO=k^{2l+1}\cot\delta_l$
diverges (approaches zero) at $\delta_l=0$ ($\pi/2$), e.g.~in the
\wave{1}{S}{0} wave at $k\approx370\;\MeV$ and \wave{3}{S}{1} wave at
$k\approx90\;\MeV$ and $400\;\MeV$~\cite{Epelbaum:2014efa}. As the qualitative
plot in Fig.~\ref{fig:qualitative} shows, the corresponding spikes may make it
more difficult to determine slopes.

\myparagraph{Fitting to a Point} A ``zero'' in
eq.~\eqref{eq:master} is induced intentionally when the observable contains a
LEC that is determined in the channel in which one tests the PC. If the
observable is tuned to exactly reproduce a certain value at some point
$(k_0,\ptyp)$, then $\calO(k_0;\Lambda_1)-\calO(k_0;\Lambda_2)=0$ -- with all
the problems mentioned just now. Obviously, one should choose the fit point to
be outside the slope-region. In the example of Sect.~\ref{sec:application},
the strength of the $3N$ interaction $H_0$ without derivatives was fixed at
each order to the $Nd$ scattering length, i.e.~using $k=0$ as fit point. That
is far away from the slope-region. At \NXLO{2}, the momentum-dependent $3N$
interaction $H_2$ was in addition determined from the triton binding energy
$B_3=8.48\;\MeV$, i.e.~the pole in the amplitude is fixed to $k_0=\sqrt{-4M
  B_3/3}\approx100\;\ii\;\MeV$. If one chooses this fit point for $H_0$ at LO
and NLO, instead of $k_0=0$, the pattern of the slopes is wiped out; see
Fig.~\ref{fig:toB3}. It appears that fitting only at $k_0$ introduces a new
low-energy scale $\ptyp$ and leaves no window $\LambdaNoPion\gg
k\gg|k_0|\approx100\;\MeV$, while the \NXLO{2} fit at both $k=0$ and $k_0$
does not suffer this limitation.

\myparagraph{Fitting in a Region} The issue is less transparent when the LEC
is not determined by exactly reproducing some data, but by least-$\chi^2$
fitting over a whole region in $k$. That is the typical case in $\mathrm{NN}$
scattering; see e.g.~Ref.~\cite{Epelbaum:2014efa}. The deviation of the fitted
result from data is more regular at any given cutoff $\Lambda$ than when it is
exactly zero at $k_0$. A pronounced spike is therefore replaced by a more
uniform behaviour inside the fit region. The comparison between two cutoffs in
eq.~\eqref{eq:master} is therefore also more uniform as a function of
$k$. Since cutoff variations can now be balanced by adjusting LECs, the
coefficients $\calC_n$ are artificially small in that r\'egime. One still
expects the cutoff dependence to decrease order-by-order, but the
characteristic slopes are harder to see since the observable is constrained by
the fit. Just like in the neighbourhood of a fit point, an observable will
first have to shed the fit constraints outside the fit region for pronounced
slopes.

Such a fit region must of course be inside the applicability range of the
EFT. Traditional fits do not take into account that the systematic
uncertainties of an EFT increase with $k$ but assign a $k$-independent
uncertainty weight. Eq.~\eqref{eq:observable} suggests that this is justified
for $k\lesssim\ptyp$ since the error varies only mildly. In that case, one can
speculate that the impact on the slopes at higher $k$ is not too big. This
limits a reasonable fit region to $k\lesssim\gamma_{t,s}$ in \EFTNoPion; and
to $k\lesssim\mpi$ in \ChiEFT. In addition, one expects clearer signals if the
same fit region is used at each order. It is difficult to see how slopes can
clearly be identified when the fit region extends far towards
$\LambdaEFT$. Practical considerations, like insufficient or low-quality data
at low momenta may well override this choice.

\myparagraph{Fitting to Pseudo-Data} As a recourse and in order
to assess the impact of a fit region on the slopes, one may create an
artificial, ``exact datum'' $\calO_0(k_0)$ at very low $k\to0$ which agrees
with low-energy data (e.g.~a scattering length, effective range, etc); and
then assess the dependence of the slope on reasonable variations of
$\calO_0(k_0)$. The goal is then not to find good agreement with actual data
at higher energies, but to test the convergence pattern.

\myparagraph{Summary: Choice of Observable} Ideal candidates for $\calO$ are
positive-definite observables which are not subject to unitarity and other
constraints, and which are nonzero and finite over a wide range in $k$ and
$\Lambda$, including the r\'egime $k\gtrsim\ptyp$ where one hopes to determine
the slope. EFT parameters/LECs should be determined at very low $k$. A good
signal may need some creativity. The choices $\calO=k^{2l+1}\cot\delta_l$,
$k^{2\pm1-2J}\bar{\delta}_{J\pm1}$ and $k^{-(2J+1)}\bar{\epsilon}_J$, with
effective-range parameters determining unknowns, appear suitable in most
scattering cases. 

%%%%%%%%%%%%%%%%%
\subsection{Miscellaneous Notes}
\label{sec:misc}

\myparagraph{Consistency Assessment vs.~``Lepage Plots''} Double-logarithmic
convergence plots are not unfamiliar. Lepage compared to data in order to
quantify how accurately the EFT reproduces experimental
information~\cite{Lepage:1997cs}. This triggered a series of influential
studies of differences between approximations and ``exact results'' in
toy-models, see e.g.~\cite{Steele:1998un, Steele:1998zc,
  Kaplan:1999qa}. Recently, Birse perused similar techniques, after removing
the strong influence of long-range Physics (One- and Two-Pion Exchange) from
empirical phase shifts in a modified effective range expansion, allowing for a
more detailed study of the residual short-distance
interactions~\cite{Birse:2007sx, Birse:2010jr}. Such investigations assume
that the correct PC is known and quantitative comparison to data is
needed.

The test advocated here aims to answer different questions: Does the
output match the assumptions? Is the theory consistent?  Recall that an EFT
may converge by itself, but not to data, if some dynamical degrees of freedom
are incorrect or missing.
For example, a \ChiEFT without dynamical $\Delta(1232)$ at $k\approx300\;\MeV$
cannot reproduce Delta resonance properties -- but it may well be consistent.
In other words, an EFT may be consistent, but not consistent with Nature.

\myparagraph{Insensitivity to Some LECs} This procedure can only help
determine if a LEC is correctly accounted for when it is needed to absorb
residual cutoff dependence.  Eq.~\eqref{eq:rge} then determines its running,
and its initial condition is fixed by some input, for example data or results
of a more fundamental theory. Some LECs do however start contributing just
because of their natural size, and not to renormalise that order. For example,
the magnetic moment of the nucleon enters the one-baryon Lagrangean of \ChiEFT
at NLO, albeit it is not needed to renormalise loops. Similarly, the
contribution of a LEC to a particular observable may be unnaturally small (or
even zero).

\myparagraph{Numerics} The analysis can be numerically indecisive. We would
trust results only if $n$ and $\LambdaEFT$ can be determined quite robustly in
a reasonably wide range to cutoffs (and, possibly, cutoff forms),
parameter sets and fit-windows. None of this provides, however, sufficient
excuse not report results.

\myparagraph{Sampling Tests} Finding that the slope at each order $Q^n$ is not
smaller than $n+1$ is necessary but not sufficient for a consistent PC. We saw
that fine-tuning, particular choices of regulator forms and observables, and
anomalously small coefficients are some reasons which may hide signals of
slopes $<n+1$ which violate the PC assumptions. If slopes are always $\ge n+1$
for a variety of independent observables, regulators etc., that may increase
confidence in PC consistency -- but cannot prove it.

%%%%%%%%%%%%%%%%%
\subsection{Outlook} 

The \ChiEFT power-counting proposals differ most starkly in the attractive
triplet partial waves of $\mathrm{NN}$ scattering since they reflect different
philosophies on how to treat the non-selfadjoint, attractive $1/r^3$ potential
at short distances which appears at leading order; see Table~\ref{tab:pc}. It
would therefore be interesting to see this test applied to the \wave{3}{P}{0}
wave and to the \wave{3}{P}{2}-\wave{3}{F}{2} system.  The test proposed here
is not necessarily a silver bullet to endorse or reject a particular counting
since its results may in the worst case be inconclusive. But that implies it
is still worth a try.

%%%%%%%%%%%%%%%%%%%%%%%%%
\acknowledgments I cordially thank the organisers of \textsc{Chiral Dynamics
  2015} in Pisa for a stimulating atmosphere, and the participants for
enlightening and entertaining discussions.  These notes grew out of the
inspirational and intense discourses at the workshops \textsc{Nuclear Forces
  from Effective Field Theory} at CEA/SPhN Saclay in 2013, and \textsc{Bound
  States and Resonances in Effective Field Theories and Lattice QCD
  Calculations} in Benasque (Spain) in 2014. I am most grateful to their
organisers and participants. Since 2013, exchanges with M.~Birse, B.~Demissie,
E.~Epelbaum, R.~Furnstahl, B.~Long, M.~Pavon Valderrama, D.~R.~Phillips,
M.~Savage, R.~G.~E.~Timmermans, U.~van Kolck and Ch.-J.~Yang allowed me to
develop these ideas into a sharper analysis tool. M.~Birse, B.~Demissie,
E.~Epelbaum, D.~R.~Phillips, and in particular the referee (DRP), suggested
important improvements to this script. I am especially indebted to ceaseless
questions by many emerging researchers. Finally, my colleagues may forgive
mistakes and omissions in referencing work and historical precedents, and
graciously continue to point out necessary corrections.
This work was supported in part by the US Department of Energy under contract
DE-FG02-95ER-40907, and by the Dean's Research Chair programme of the
Columbian College of Arts and Sciences of The George Washington University.

\end{document}